\documentclass[a4paper,11pt]{article}
\usepackage[DIV10]{typearea}
\usepackage[latin1]{inputenc}
\usepackage[T1]{fontenc}
\usepackage{graphicx}
\usepackage[english]{babel}
\usepackage{amssymb}
\usepackage{color}
\usepackage{xspace}
\usepackage{textcomp}
\usepackage{hyperref}
\usepackage{epic}
\usepackage{eepic}
\usepackage{amsmath,bm}
\usepackage[footnotesize]{caption2}

\DeclareMathSymbol{\Gamma}{\mathalpha}{letters}{"00}
\DeclareMathSymbol{\Delta}{\mathalpha}{letters}{"01}
\DeclareMathSymbol{\Theta}{\mathalpha}{letters}{"02}
\DeclareMathSymbol{\Lambda}{\mathalpha}{letters}{"03}
\DeclareMathSymbol{\Xi}{\mathalpha}{letters}{"04}
\DeclareMathSymbol{\Pi}{\mathalpha}{letters}{"05}
\DeclareMathSymbol{\Sigma}{\mathalpha}{letters}{"06}
\DeclareMathSymbol{\Upsilon}{\mathalpha}{letters}{"07}
\DeclareMathSymbol{\Phi}{\mathalpha}{letters}{"08}
\DeclareMathSymbol{\Psi}{\mathalpha}{letters}{"09}
\DeclareMathSymbol{\Omega}{\mathalpha}{letters}{"0A}
\DeclareMathSymbol{\varGamma}{\mathalpha}{operators}{"00}
\DeclareMathSymbol{\varDelta}{\mathalpha}{operators}{"01}
\DeclareMathSymbol{\varTheta}{\mathalpha}{operators}{"02}
\DeclareMathSymbol{\varLambda}{\mathalpha}{operators}{"03}
\DeclareMathSymbol{\varXi}{\mathalpha}{operators}{"04}
\DeclareMathSymbol{\varPi}{\mathalpha}{operators}{"05}
\DeclareMathSymbol{\varSigma}{\mathalpha}{operators}{"06}
\DeclareMathSymbol{\varUpsilon}{\mathalpha}{operators}{"07}
\DeclareMathSymbol{\varPhi}{\mathalpha}{operators}{"08}
\DeclareMathSymbol{\varPsi}{\mathalpha}{operators}{"09}
\DeclareMathSymbol{\varOmega}{\mathalpha}{operators}{"0A}

\newcommand{\D}{\mathrm{d}}
\newcommand{\E}{\mathrm{e}}

\def\beq{\begin{equation}}
\def\eeq{\end{equation}}
\def\bea{\begin{eqnarray*}}
\def\eea{\end{eqnarray*}}

\newcommand{\stau}{{\widetilde{\tau}}}		

\newcommand{\mstau}{m_{\stau_1}}    
\newcommand{\pt}{p_\text{T}}				
\newcommand{\thest}{\theta_{\stau}}
\newcommand{\R}{{r_{\beta}}}	
\newcommand{\IL}{\mathcal{L}_{{}^{\int}}}
\newcommand{\GEV}{\ensuremath{\,\textnormal{GeV}}}
\newcommand{\TEV}{\ensuremath{\,\textnormal{TeV}}}
\newcommand{\SEC}{\ensuremath{\,\textnormal{s}}}
\newcommand{\pb}{\ensuremath{\,\textnormal{pb}}}
\newcommand{\fb}{\ensuremath{\,\textnormal{fb}}}

\newcommand{\SY}[1]{\widetilde{#1}}
\definecolor{evgray}{gray}{.48}
\begin{document}

\date{\mbox{ }}

\title{ 
{\normalsize  
3rd June 2011 \hfill\mbox{}\\}
\vspace{2cm}
\bf
Production of long-lived staus in the Drell-Yan process\\[8mm]}
\author{Jan Heisig and J\"{o}rn Kersten\\[2mm]
{\small\it II.~Institute for Theoretical Physics, University of Hamburg,
Germany}\\
{\small\tt jan.heisig@desy.de, joern.kersten@desy.de}
}

\maketitle

\thispagestyle{empty}

\vspace{1cm}

\begin{abstract}
We investigate the phenomenology of the gravitino dark matter scenario
with a stau as the next-to-lightest supersymmetric particle at the
LHC\@.  For a wide range of gravitino masses the lighter stau is stable
on the scale of a detector and gives rise to a prominent signature as a
``slow muon''. 
The direct stau production via the Drell-Yan process is always present
and independent of the mass spectrum of the other superparticles, thus
providing a lower bound for the discovery potential of this scenario.
Performing a careful analysis with particular emphasis on the criteria
for observing stau pairs and for distinguishing them from the
background, we find that the $14\TEV$ run of the LHC 
has a promising potential for finding long-lived staus from Drell-Yan
production up to very large stau masses.
\end{abstract}

\clearpage

%===========================================================================================
\section{Introduction}
%===========================================================================================

Supersymmetry (SUSY) with conserved R-parity and a gravitino as the
lightest superparticle (LSP) is a viable alternative to the most widely
studied scenario with a neutralino LSP\@.  A stable gravitino is a
perfectly good dark matter candidate \cite{Fayet:1981sq,Pagels:1981ke} and may even
be regarded as favored, since it alleviates the cosmological gravitino
problem, allowing for a higher reheating temperature after
inflation~\cite{Ellis:1984er,Bolz:1998ek}.
As the superpartner of the graviton, the gravitino takes part only in
the gravitational interaction.  Therefore, the next-to-LSP (NLSP) is
typically quite long-lived.%
\footnote{The same is true in scenarios with an axino LSP, whose
 interactions are strongly suppressed by the large Peccei-Quinn scale,
 see, e.g., \cite{Brandenburg:2005he}.  The NLSP can also be long-lived
 if its mass is very close to that of a neutralino
 LSP~\cite{Jittoh:2005pq}.  We do not study these alternatives in detail
 but expect the same results as in the case considered.}
For a charged NLSP, this leads to a spectacular signature at colliders,
charged tracks leaving the detector and no missing transverse energy.
It could even be possible to capture NLSPs and to study their decays in
detail, thus measuring the strength of their coupling to the LSP and
the LSP's spin \cite{Buchmuller:2004rq}.  In this way, observations of
the NLSP could lead to an indirect determination of the nature of the LSP.

In this work, we consider a charged slepton NLSP\@.  In the following,
we refer to the lightest charged slepton as the stau $\stau_1$ and allow
for mixing between $\stau_\text{R}$ and $\stau_\text{L}$, the
superpartners of the right- and left-handed tau, respectively.  Of
course, the results are also valid for a selectron or smuon NLSP\@.
For a wide range of gravitino and stau masses, the stau NLSP lifetime
\begin{equation} \label{eq:taustau}
\tau_\stau\simeq6\times10^{4}\,{\rm s}\, \left(\frac{m_{3/2}}{\!\GEV}\right)^2\left(\frac{m_{\stau}}{100\GEV}\right)^{-5}
\end{equation}
is larger than about $10^{-7}\SEC$.
Then the stau is metastable, i.e., it usually leaves an LHC detector
before decaying.

In this scenario
catalyzed big bang nucleosynthesis \cite{Pospelov:2006sc} leads to an
upper bound of roughly $10^4\SEC$ on the stau lifetime. 
While this bound can in principle be satisfied by lowering the gravitino
mass sufficiently, a short lifetime is also obtained for gravitino
masses in the GeV range and a relatively heavy stau
\cite{Pradler:2007is,Kersten:2007ab} due to the dependence of
$\tau_\stau$ on $\mstau^{-5}$.

Previous studies
of the LHC phenomenology of metastable staus have concentrated on the
production via decays of heavier superparticles, assuming specific
scenarios for SUSY breaking
\cite{Nisati:1997gb,Hinchliffe:1998ys,Ambrosanio:2000ik,Ambrosanio:2000zu,Sjolin:2002fx,Feng:2004mt,DeRoeck:2005bw,Ellis:2006vu,Feng:2009yq,Feng:2009bd,Chen:2009gu,Ito:2009xy,Heckman:2010xz,Kitano:2010tt,Endo:2010ya,Endo:2011uw}.
See also \cite{Fairbairn:2006gg} for a comprehensive review of the topic.
Here we do not restrict ourselves to a specific SUSY-breaking scenario
and we focus on the direct production of staus via the neutral current 
Drell-Yan (DY) process, which possesses interesting properties:
\begin{itemize}
\item
The DY contribution is independent of all MSSM parameters except
$\mstau$ and the stau mixing angle $\thest$, enabling a
model-independent analysis.
\item
The DY process is always present.
Together with the previous point, conservatively
this leads to an assured
discovery potential and strict exclusion limits in a class of scenarios
characterized only by $\mstau$ and $\thest$.
\end{itemize}
Thus, it is natural to ask for the required luminosity which provides this 
robust exclusion limit (and the discovery reach) in
the $7\TEV$ and $14\TEV$ LHC run.
Although the DY production of staus has been included in some studies 
\cite{Rajaraman:2006mr,Rajaraman:2007ae,Endo:2011uw},
the focus of these works has been different 
and---to our knowledge---this question has only been addressed in a brief remark 
in the review \cite{Raklev:2009mg}, where the aim is only 
a rough estimate with fairly conservative assumptions.
Here, we perform a careful analysis, in particular examining the
dependence on the imposed cuts and using the proper statistics for small
event numbers.  We also take into account the latest information
from the LHC experiments on discriminating heavy stable charged
particles from muons.  We find that 
the opposite-sign stau pair from DY production
allows for a clean signal region up to
very large integrated luminosities. Thus, in spite of its small cross
section the DY process is able to provide an interesting
discovery and exclusion potential, even for relatively large stau
masses and significantly better than estimated in \cite{Raklev:2009mg}.

The paper is organized as follows. In section~\ref{sec:DrellYan} we
examine the DY process, discussing the criteria for
observing stau pairs and for distinguishing them from the background.
This allows us to derive the LHC's discovery reach and exclusion
potential in section~\ref{sec:Discovery}. 
In section \ref{sec:production} we briefly compare DY production 
with the production of staus from the decay of heavier superparticles. 
The purpose of this consideration is to
estimate for which SUSY spectra the exclusion limit
from direct DY production is tight
and for which spectra it tends to be overly loose.
We will see that the direct DY production can be dominant 
for large mass gaps between the stau NLSP and the colored superparticles.
Unless noted otherwise, we discuss the $14\TEV$ LHC run in what follows.

%===========================================================================================
\section{Staus from the Drell-Yan process} \label{sec:DrellYan}
%===========================================================================================

Let us first discuss in detail the DY production of staus, the
expected signal in a detector at the LHC and the background suppression.
All events have been generated with \textsc{MadGraph/MadEvent
4} \cite{alwall-2007} and its \textsc{Pythia} \cite{sjostrand-2006-0605}
interface. For the event generation in \textsc{MadEvent} and 
\textsc{Pythia} we enabled the MLM matching scheme \cite{alwall-2008-53}.  
The \textsc{Cteq6l1} PDF set \cite{Pumplin:2002vw} has been used.

%-------------------------------------------------------------------------------------------
\subsection{Background}\label{sec:background}
%-------------------------------------------------------------------------------------------

Staus as heavy metastable charged particles usually leave the detector.
This leads to a signal in the tracker and muon chambers. Muons are the
only background.  Therefore, we first study this background in order to
devise suitable cuts, which we can then take into account in the
calculation of the signal in the next subsection.

The di-muon rate for the moderate and high $\pt$-range is much smaller
than the single muon rate and the possible sources are considerably
fewer in the case of di-muons.
For a $p_{\rm T}$-cut smaller and around
$15\GEV$, $b$- and $c$-decays are the dominant sources of di-mouns.
 Above $15\GEV$ the DY production begins to dominate \cite{Albajar:687291}.
In addition to a high $p_{\rm T}$-cut, the $b$- and $c$-contributions can be further reduced 
by isolation cuts. This feature relies on the fact that $b$ and $c$ quarks are 
always produced close to jets while muons from heavy mother particles 
(like $Z$ or $t$) tend to be well-separated from the other decay 
products---they are isolated \cite{Amapane:687467}.

As preselection cuts on the data, we require
two opposite-sign muon-like particles each satisfying
\begin{itemize}
 \item $p_{\rm T}$-cut: $p_{\rm T}>50\GEV$
 \item Barrel cut on the pseudorapidity: $|\eta|<2.5$
 \item Isolation cut: $\Delta\mathcal{R}_{\mu, {\rm jet}}=\sqrt{\Delta\eta_{\mu, {\rm jet}}^2+\Delta\phi_{\mu, {\rm jet}}^2}>0.5$ for jets with $p_{\rm T}>50\GEV$.
\end{itemize}
With the $p_{\rm T}$- and the isolation cut, a sufficient rejection
of the $b$- and $c$-contributions should easily be obtained. This is why we
refrain from running a detector simulation for this issue. All the same
we made no effort to specify the isolation algorithm. Instead we apply the respective 
cuts directly on the remaining background (and signal) at the level of the 
\textsc{Pythia} Les Houches Event output.
The $p_{\rm T}$-cut also rejects muon pairs from on-shell $Z$ decay.
Therefore, an additional cut on the invariant mass of the muon pair,
which was used in \cite{Rajaraman:2006mr}, would not have a significant
effect.

As the remaining background, we consider di-muons from the DY process
and from $t\bar{t}$-production.  We include contributions up to order 
$\alpha_{\rm s}^2$ (two jets) in the case of the DY process and 
$\alpha_{\rm s}^3$ (three jets) in the case of $t\bar{t}$-production and generate 
$2\times 10^{5}$ unweighted events for the analysis. 
The total normalization of the cross section was fixed from the leading order DY 
di-muon production (without jets) from MadEvent multiplied by a constant
$K$-factor, conservatively chosen to be $1.4$, to account for
next-to-leading order (NLO) corrections.
The total di-muon cross section after applying the above preselection cuts is then $\sigma_{\rm B}\simeq 25\pb$.

%=====================
%    \                                           |
%      \                                         |
%        \                                       |
\begin{figure}[tb]
\centering
\setlength{\unitlength}{1\textwidth}
\begin{picture}(0.95,0.53)
  \put(0.51,0.03){\includegraphics[scale=1.193]{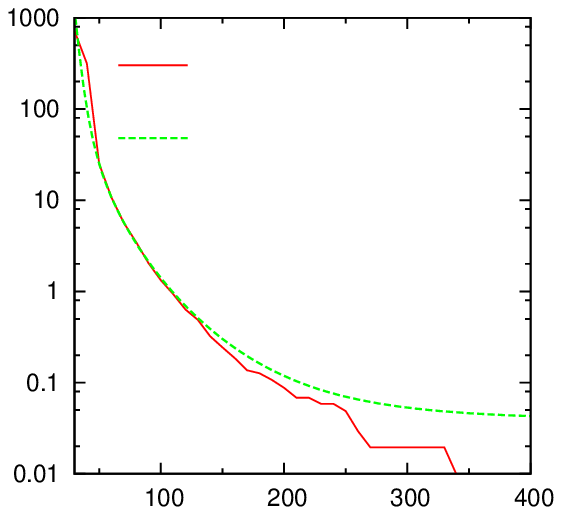}}
    \put(0.49,0.219){\rotatebox{90}{$\sigma_{\rm B}\;[$pb$]$}}
    \put(0.697,0.0){$\pt^\text{min}\;[$GeV$]$}
    \put(-0.013,-0.0093){
    \put(0.7,0.4122){\footnotesize generated via MadEvent/}
    \put(0.7,0.385){\footnotesize Pythia}
    \put(0.7,0.352){\footnotesize conservative fit for }
    \put(0.7,0.327){\footnotesize $\pt>50\GEV$}
    }
  \put(-0.002,0.03){\includegraphics[scale=1.193]{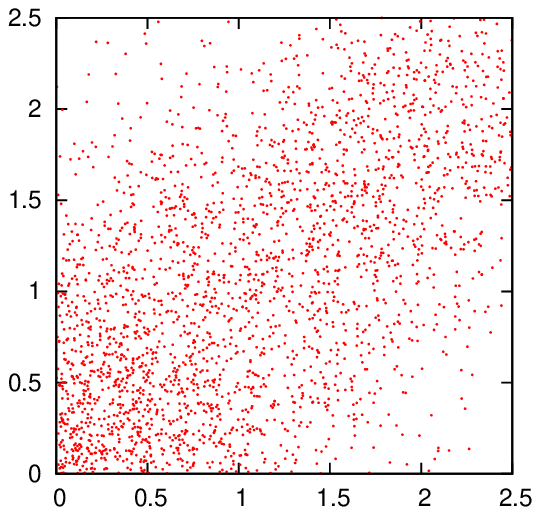}}
    \put(-0.027,0.235){\rotatebox{90}{$|\eta^+|$}}
    \put(0.227,0.0){$|\eta^-|$}
\end{picture}
\caption{Left: Pseudorapidity distribution of the considered di-muon
background after applying the preselection cuts, where $\eta^\pm$ denotes the pseudorapidity of $\mu^\pm$. Right: Total di-muon
cross section after applying the preselection cuts but as a function of
the variable $\pt$-cut discussed at the end of section \ref{sec:Discovery} (again on both muons).
Both plots are valid for the $14\TEV$ LHC.}
\label{fig:dimu}
\end{figure}
%                                      \         |
%                                        \       |
%                                          \     |
%=====================

\subsubsection*{Velocity measurement}

The crucial tool for distinguishing muons from staus is the velocity measurement. 
A significant fraction of staus with a mass of several hundred GeV will
have a velocity well below the speed of light, whereas the background
muons are always ultra-relativistic. 
Unfortunately, measuring the velocity is much more involved than
measuring, for example, the momentum. Therefore, the experimental
uncertainty is much larger, and a cut on the velocity will not reject all background muons.

There are two distinct ways of measuring the velocity of muon-like
particles, the ionization energy loss ($\D E/\D x$) inside the tracker
and the time-of-flight (ToF) measurement, which measures the time
between the bunch crossing and the passing through the muon chambers.
The former measurement only provides information up to $\beta \lesssim 0.9$
while the latter is mainly limited from below---at the design luminosity
of the LHC, particles with velocities less than about $0.6$ cannot be
assigned to the correct bunch crossing anymore.

Although the precision of each velocity measurement is not overwhelming,
a combination of both measurements provides a highly efficient
background rejection. This is due to the fact that for stau signal
events these two measurements are clearly correlated while for
background events no correlation is present \cite{PoS2008LHC012}.
According to \cite{CMS-PAS-EXO-08-003}
a cut using both measurements,
 $\beta_{\D E/\D x}$, $\beta_{\rm ToF}<0.8$,
leads to a background rejection factor of about $10^{-7}$ for single stau candidates. 

Thus, if the probability of a mis-identification
of two muons within one event is not correlated,
a background rejection factor of $\R=10^{-14}$ could 
be achievable.%
\footnote{A possible source of such a correlation would be the presence of a highly correlated 
$\eta^+$-$\eta^-$-distribution together with a strong dependence of the velocity resolution function on $\eta$. 
However, figure \ref{fig:dimu} shows that already the former is not the case.}
However, for $\R=10^{-14}$, the relevant background appears only at very high
luminosities, when pile-up from different bunch crossings becomes relevant.
This fact might lower the background rejection with respect to the
na\"{\i}ve expectation of $\R=10^{-14}$.  To our knowledge, currently
no quantitative study exists about this issue.
In any case, we shall
see later that a sufficient background rejection
(which enables a three-event exclusion
over the whole considered region for $\mstau$)
can already be achieved with $\R\simeq10^{-10}$.
Thus, using this value is both sufficient and conservative. 
To show the effect of a looser background rejection 
we will also display the results under the pessimistic 
assumption that we do not gain anything by requiring
\emph{two} stau candidates, hence applying $\R=10^{-7}$. 

Following these considerations, we will apply the cuts $0.6 < \beta < 0.8$
on the signal events in order to ensure both a working ToF measurement
and a sufficient background rejection.

%-------------------------------------------------------------------------------------------
\subsection{Signal}\label{sec:staumixing}
%-------------------------------------------------------------------------------------------

The direct stau production via DY only involves two parameters of the
more than 100 free MSSM parameters, the stau mass $\mstau$ and the
mixing angle $\thest$.

%=====================
%    \                                           |
%      \                                         |
%        \                                       |
\begin{figure}[bt]
\centering
\setlength{\unitlength}{1\textwidth}
\begin{picture}(0.91,0.45)
  \put(0.0,0.042){\includegraphics[scale=1.32]{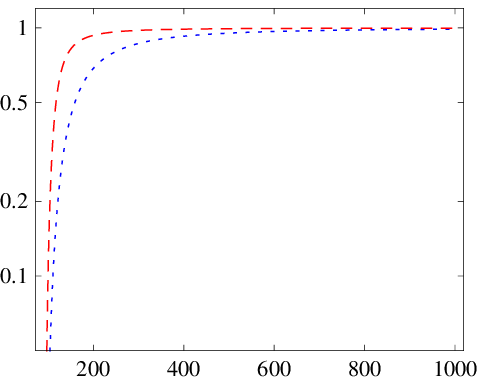}}
    \put(-0.045,0.185){\rotatebox{90}{$\sigma_{\rm e}/\sigma_{\rm ew}$}}
    \put(0.188,0.0){$\sqrt{\hat{s}}\;[$GeV$]$}
   \put(0.5,0.04){\includegraphics[scale=1.32]{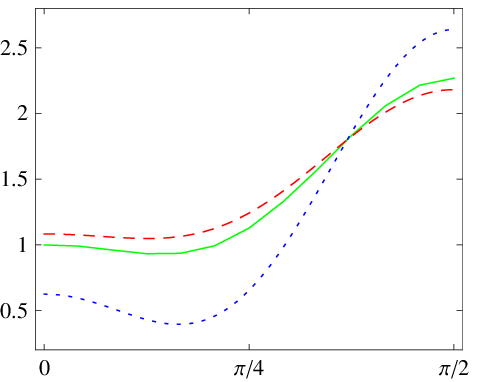}}
     \put(0.46,0.165){\rotatebox{90}{$\sigma_{\rm ew}\;$[a.u.]}}
     \put(0.715,0.0){$\thest$}
\end{picture}
\caption{Left:~Ratio between the electromagnetic contribution and the complete
electroweak cross section (\ref{eq:DYcross-section}) for $\thest=0$ as
a function of $\sqrt{\hat{s}}$ for up-type quarks (red dashed line) and
down-type quarks (blue dotted line). In each case the ratio is normalized to be one at $\hat{s}\to\infty$.
Right:~Cross section for direct di-stau production
$pp\to Z,\gamma\to \stau_1^+\stau_1^-$ as a function of the stau mixing
angle. The curves are obtained from a MadEvent simulation with
$\mstau=500\GEV$ (solid green, normalized to be one at $\thest=0$) as
well as directly from (\ref{eq:DYcross-section}) for up-type quarks (red
dashed line) and down-type quarks (blue dotted line) for
a center-of-mass energy
$\sqrt{\hat{s}}=1000\GEV$. The two latter curves are absolutely
normalized to be equal to the MadEvent prediction at their intersection
point.}
\label{fig:theta-stau}
\end{figure}
%                                      \         |
%                                        \       |
%                                          \     |
%=====================

\subsubsection*{Dependence on the mixing angle}

We define the stau mixing matrix via
\beq
  \label{}
 \begin{pmatrix} \tilde{\tau}_1 \\
\tilde{\tau}_2
\end{pmatrix}
={\cal M}^{\tilde{\tau}}
 \begin{pmatrix} \tilde{\tau}_{\text{R}} \\
\tilde{\tau}_{\text{L}}
\end{pmatrix}
=
 \begin{pmatrix} \cos\theta_{\tilde{\tau}}& \sin\theta_{\tilde{\tau}}\\
-\sin\theta_{\tilde{\tau}}&\cos\theta_{\tilde{\tau}} 
\end{pmatrix}
 \begin{pmatrix} \tilde{\tau}_{\text{R}} \\
\tilde{\tau}_{\text{L}}
\end{pmatrix}
\label{eq:mixing-def}
\eeq
and $m_{\stau_1}\leq m_{\stau_2}$.
The dependence of the di-stau cross section on the mixing angle $\thest$
can be discussed by considering the tree-level parton-level cross section for
$q\bar{q} \to Z,\gamma \to \stau_i^+\stau_j^-$,
\beq
  \left(  \frac{\D\hat{\sigma}}{\D\hat{t}}\right)^{q}_{ij}=
    \frac{e^4}{8\pi}\, \frac{\hat{u}\hat{t}-m_{\stau_i}^2m_{\stau_j}^2}{\hat{s}^4}
    \left[
      Q_{q}^2 \delta_{ij}
      +({g_{\text{V}}^{q}}^2+{g_{\text{A}}^{q}}^2)\frac{g_{\stau_i\stau_j}^2}{(1-M_{{Z}}^2/\hat{s})^2}
      -\frac{2Q_{q}\,g_{\text{V}}^{q} \,\delta_{ij}\,g_{\stau_i\stau_j}}{1-M_{{Z}}^2/\hat{s}}
    \right]\,,
  \label{eq:DYcross-section}
\eeq
where
\beq
  g_{\stau_i\stau_j}=
    \frac{1}{c_{\text{W}}s_{\text{W}}}\left[\left(-\frac{1}{2}+s_{\text{W}}^2\right){\cal M}_{i\text{L}}^\stau {\cal M}_{j\text{L}}^\stau +s_{\text{W}}^2 {\cal M}_{i\text{R}}^\stau {\cal M}_{j\text{R}}^\stau\right]
  \label{eq:g-stau-def}
\eeq
and
\beq
    g_{\text{V}}^{q}=\frac{T^3_{q}-2Q_{q}s_{\text{W}}^2}{2c_{\text{W}}s_{\text{W}}}\,,
    \quad g_{\text{A}}^{q}=\frac{T^3_{q}}{2c_{\text{W}}s_{\text{W}}}\,.
  \label{eq:g-quark-def}
\eeq
Here $q$ denotes the flavor of the annihilating quark, while $Q_q$ and
$T^3_q$ are its electric charge and the third component of its weak
isospin, respectively.
Besides, $c_{\text{W}}\equiv\cos{\theta_W}$, $s_{\text{W}}\equiv\sin{\theta_{\text{W}}}$, and
$\hat{s}$, $\hat{t}$, $\hat{u}$ are the Mandelstam variables.
(Taking into account the width $\Gamma_Z\ll M_Z$ is not vital for the following argumentation.) 

A change in $\thest$ has an impact on $g_{\stau_i\stau_j}$ only. Thus,
it alters the ratio between the three terms in square brackets in
\eqref{eq:DYcross-section}, which are the electromagnetic, the weak
and the interference terms, respectively.
This change is almost independent of the kinematics. The terms in square
brackets contain $\hat{s}$ as the only kinematic variable, and even this
dependence becomes negligible when exceeding a few times $M_{{Z}}$. The
left panel of figure~\ref{fig:theta-stau} shows the ratio between the
first term and all terms in (\ref{eq:DYcross-section}) as a function of
$\hat{s}$ (arbitrarily normalized). From a few hundred GeV on the ratio
is almost constant. Thus, in this region a change in $\thest$ will only
shift the overall cross section 
without any impact on the kinematic distributions.
In fact, the dependence of the cross section on
$\thest$ shown in the right panel of figure \ref{fig:theta-stau} is
applicable to all stau masses considered in the following.

For the plots in figures \ref{fig:sigma-mass} and \ref{fig:distau-obs} 
we consider the case $\thest=0$. 
However, the minimum of
$\sigma_{\rm ew}(\thest)$ is at $\thest^{\rm min}\neq0$ and it is
about $7\%$ lower than the value at $\sigma_{\rm ew}(0)$. Therefore,
when estimating the discovery potential and exclusion limits we
conservatively choose $\thest=\thest^{\rm min}$. 
The limits for other values of $\thest$ can easily be obtained 
from the displayed curves. Since it turns out that we can achieve 
a (almost) clean signal region, the required luminosity is, to a very good
approximation, simply proportional to the inverse of the cross section.

\subsubsection*{Dependence on the stau mass}
 
To show the dependence on the stau mass $\mstau$ we simulated the DY
production in $21$ mass steps from $100\GEV$ to $1000\GEV$, generating
$5\times10^{4}$ events for each one and again considering diagrams up to
order $\alpha_{\rm s}^2$ (two jets). We obtained the normalization
of the total cross section from the corresponding leading-order
computation (without jets) from MadEvent, corrected by a constant
$K$-factor of $1.35$ \cite{PhysRevD.57.5871}.  This value was found
for $\mstau \gtrsim 200\GEV$ considering NLO QCD corrections.
Additionally including SUSY QCD contributions at NLO and 
next-to-leading logarithmic accuracy yields a $K$-factor between
roughly $1.29$ and $1.36$
\cite{1999PhRvL..83.3780B,Bozzi:2007qr,Bozzi:2007tea}.

%=====================
%    \                                           |
%      \                                         |
%        \                                       |
\begin{figure}[tb]
\centering
\setlength{\unitlength}{1\textwidth}
\begin{picture}(0.65,0.52)
  \put(0.0,0.02){\includegraphics[scale=0.64, bb= 20 0 660 330, clip=true]{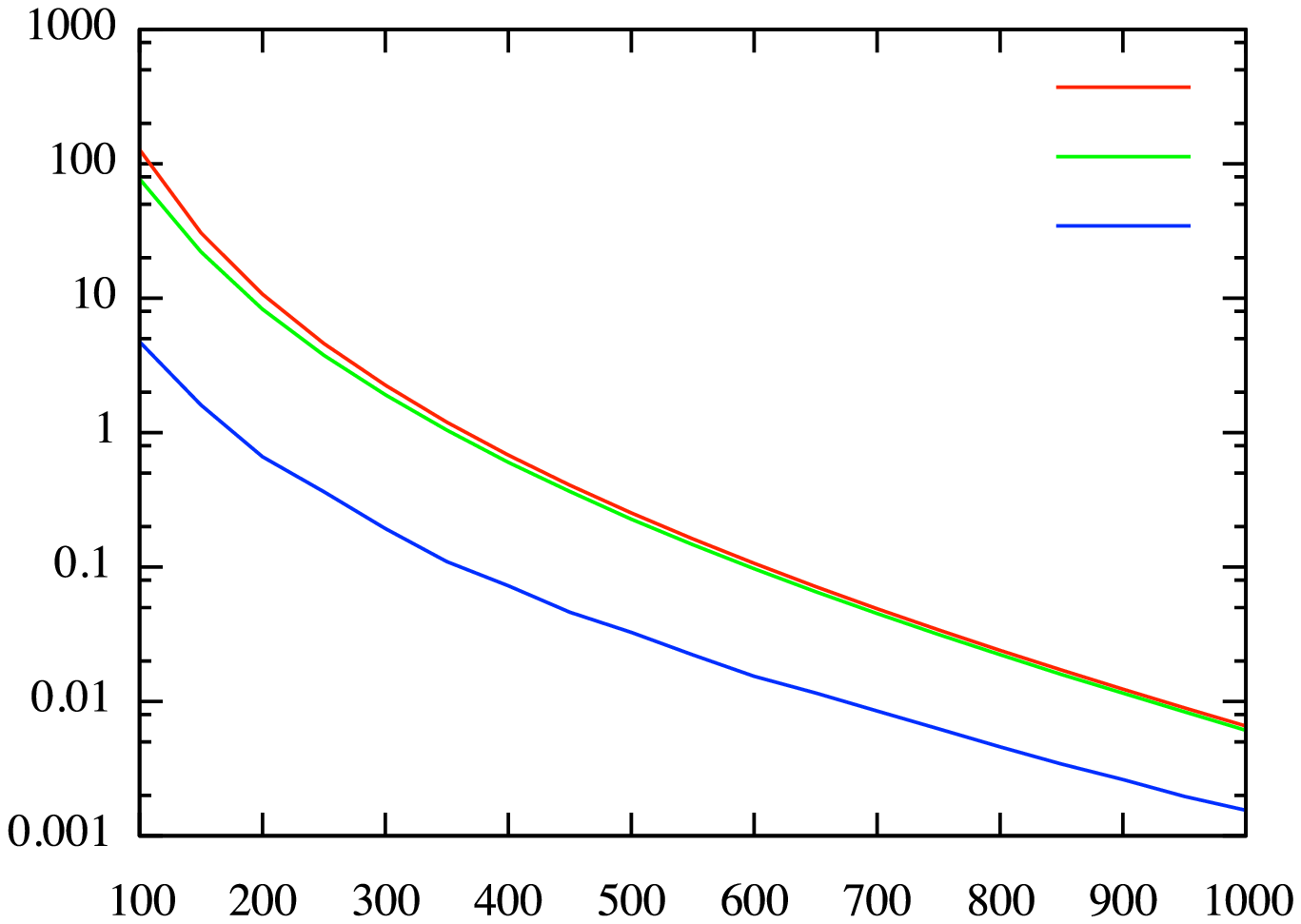}}
    \put(-0.025,0.19){\rotatebox{90}{cross section $[$fb$]$}}
    \put(0.283,0.0){$m_{\stau_1}\;[$GeV$]$}
       \put(0.375,0.429){\footnotesize without cuts}
       \put(0.3368,0.3963){\footnotesize preselection cuts}
       \put(0.212,0.3636){\footnotesize preselection and velocity cuts}
\end{picture}
\caption{Cross section for direct di-stau production $pp\to Z,\gamma\to \stau_1^+\stau_1^-$ for $\thest=0$ ($\stau_1=\stau_{\text{R}}$) 
as a function of
$m_\stau$ at the $14\TEV$ LHC\@.
The impact of the preselection cuts 
($|\eta|<2.5$, $\pt >50\GEV$, $\Delta\mathcal{R}_{\stau, {\rm jet}}>0.5$)
and the velocity cut ($0.6<\beta<0.8$) is displayed.
}
\label{fig:sigma-mass}
\end{figure}
%                                      \         |
%                                        \       |
%                                          \     |
%=====================

\begin{figure}[tbhp]
\centering
\setlength{\unitlength}{1\textwidth}
\begin{picture}(0.95,0.93)
\linethickness{0.3mm}
 \put(0.0214166666666667,0.480133333333333){\includegraphics[scale=0.60546, bb= 57 0 460 300, clip=true]{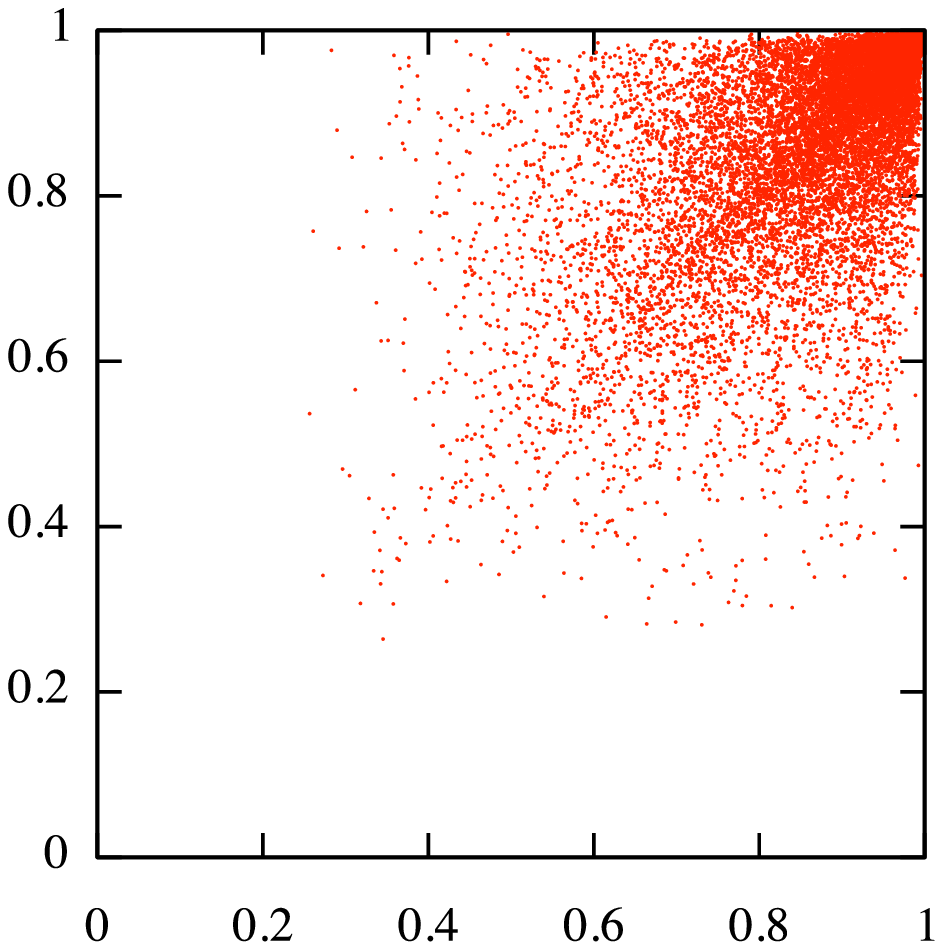}}
    \put(0.0,0.708318){\rotatebox{90}{$\beta^+$}}
    \put(0.259482666666666, 0.473301333333333){$\beta^-$}
    \put(0.107538666666667,0.559916666666666){\footnotesize{$14342$ events}}
    \put(0.0967146666666666, 0.818533333333333){\color{green}\line(1,0){0.205824}}
    \put(0.302538666666666, 0.818533333333333){\color{green}\line(0,1){0.068608}} 
    \put(0.0967146666666666, 0.749389333333333){\color{green}\line(1,0){0.205824}}
    \put(0.302538666666666, 0.749389333333333){\color{green}\line(0,-1){0.205824}}  
    \put(0.371682666666666, 0.818533333333333){\color{green}\line(1,0){0.068608}}
    \put(0.371682666666666, 0.818533333333333){\color{green}\line(0,1){0.068608}} 
    \put(0.371682666666666, 0.749389333333333){\color{green}\line(0,-1){0.205824}} 
    \put(0.371682666666666, 0.749389333333333){\color{green}\line(1,0){0.068608}} 
    \put(0.302538666666666, 0.818533333333333){\color{black}\line(0,-1){0.0159728}}
    \put(0.302538666666666, 0.749389333333333){\color{black}\line(0,1){0.0159728}}
    \put(0.302538666666666, 0.775653333333333){\color{black}\line(0,1){0.0160264}}
    \put(0.371682666666666, 0.818533333333333){\color{black}\line(0,-1){0.0159728}}
    \put(0.371682666666666, 0.749389333333333){\color{black}\line(0,1){0.0159728}}
    \put(0.371682666666666, 0.775653333333333){\color{black}\line(0,1){0.0160264}}
    \put(0.371682666666666, 0.749389333333333){\color{black}\line(-1,0){0.0159728}}
    \put(0.302538666666666, 0.749389333333333){\color{black}\line(1,0){0.0159728}}
    \put(0.328802666666666, 0.749389333333333){\color{black}\line(1,0){0.01608}}
    \put(0.371682666666666, 0.818533333333333){\color{black}\line(-1,0){0.0159728}}
    \put(0.302538666666666, 0.818533333333333){\color{black}\line(1,0){0.0159728}}
    \put(0.328802666666666, 0.818533333333333){\color{black}\line(1,0){0.01608}}
 \put(0.510583333333333,0.480133333333333){\includegraphics[scale=0.60546, bb= 57 0 460 300, clip=true]{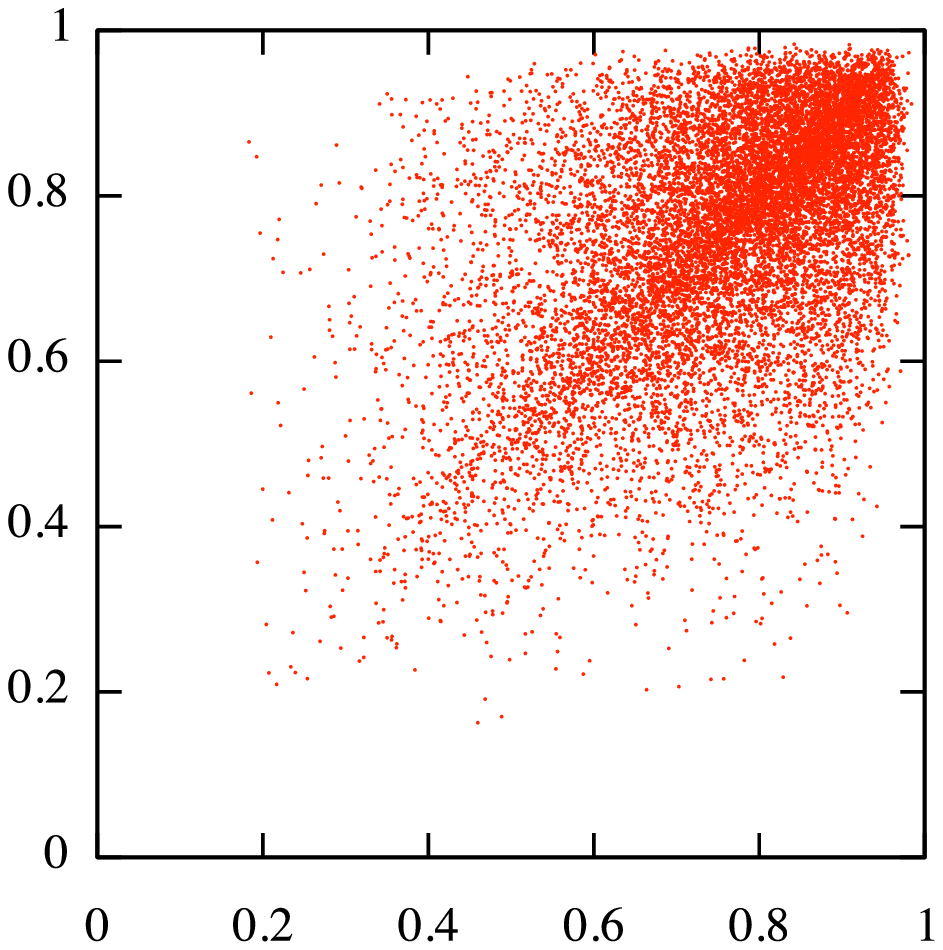}}
    \put(0.488333333333333,0.708318){\rotatebox{90}{$\beta^+$}}
    \put(0.748649333333333, 0.473301333333333){$\beta^-$}
    \put(0.593705333333333,0.559916666666666){\footnotesize{$13709$ events}}
    \put(0.585881333333333, 0.818533333333333){\color{green}\line(1,0){0.205824}}
    \put(0.791705333333333, 0.818533333333333){\color{green}\line(0,1){0.068608}} 
    \put(0.585881333333333, 0.749389333333333){\color{green}\line(1,0){0.205824}}
    \put(0.791705333333333, 0.749389333333333){\color{green}\line(0,-1){0.205824}}  
    \put(0.860849333333333, 0.818533333333333){\color{green}\line(1,0){0.068608}}
    \put(0.860849333333333, 0.818533333333333){\color{green}\line(0,1){0.068608}} 
    \put(0.860849333333333, 0.749389333333333){\color{green}\line(0,-1){0.205824}} 
    \put(0.860849333333333, 0.749389333333333){\color{green}\line(1,0){0.068608}} 
    \put(0.791705333333333, 0.818533333333333){\color{black}\line(0,-1){0.0159728}}
    \put(0.791705333333333, 0.749389333333333){\color{black}\line(0,1){0.0159728}}
    \put(0.791705333333333, 0.775653333333333){\color{black}\line(0,1){0.0160264}}
    \put(0.860849333333333, 0.818533333333333){\color{black}\line(0,-1){0.0159728}}
    \put(0.860849333333333, 0.749389333333333){\color{black}\line(0,1){0.0159728}}
    \put(0.860849333333333, 0.775653333333333){\color{black}\line(0,1){0.0160264}}
    \put(0.860849333333333, 0.749389333333333){\color{black}\line(-1,0){0.0159728}}
    \put(0.791705333333333, 0.749389333333333){\color{black}\line(1,0){0.0159728}}
    \put(0.817969333333333, 0.749389333333333){\color{black}\line(1,0){0.01608}}
    \put(0.860849333333333, 0.818533333333333){\color{black}\line(-1,0){0.0159728}}
    \put(0.791705333333333, 0.818533333333333){\color{black}\line(1,0){0.0159728}}
    \put(0.817969333333333, 0.818533333333333){\color{black}\line(1,0){0.01608}}
  \put(0.0258333333333333,0.0148){\includegraphics[scale=0.60546, bb= 60 0 480 300, clip=true]{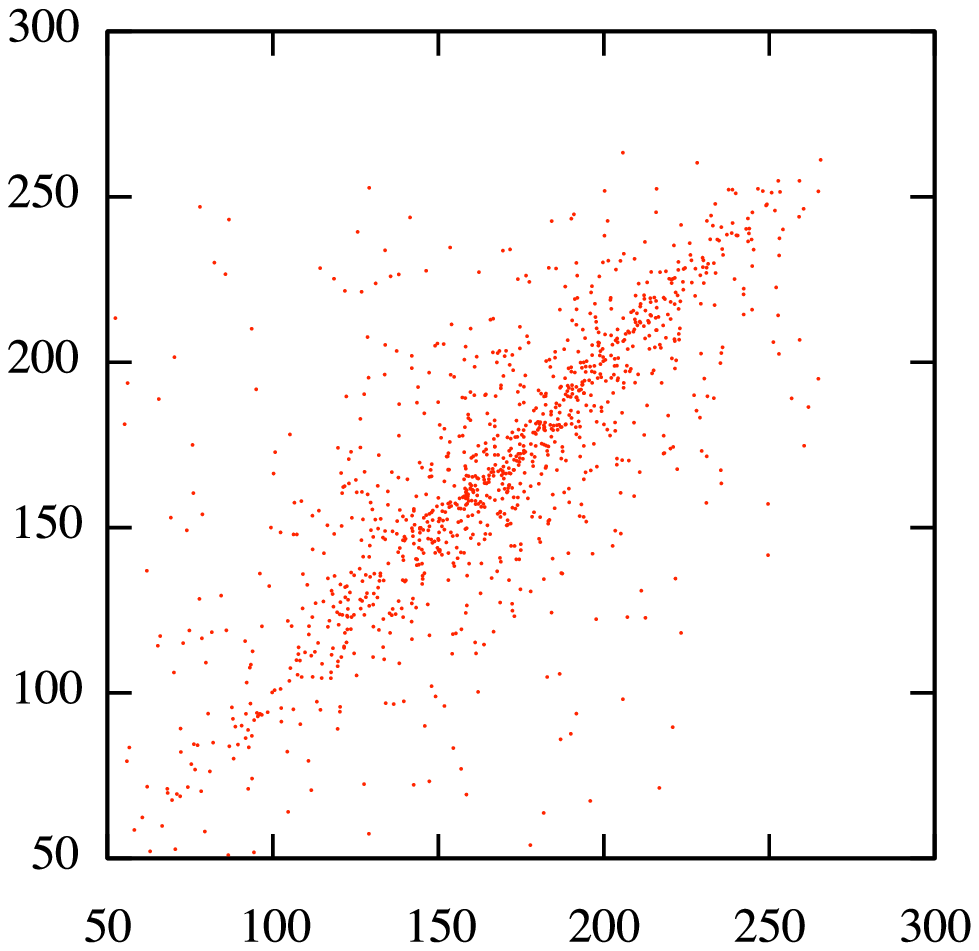}}
    \put(0.0,0.206333333333333){\rotatebox{90}{$\pt^+\;[$GeV$]$}}
    \put(0.233333333333333,0.0){$\pt^-\;[$GeV$]$}
  \put(0.53216,0.0148){\includegraphics[scale=0.60546, bb= 76 0 480 300, clip=true]{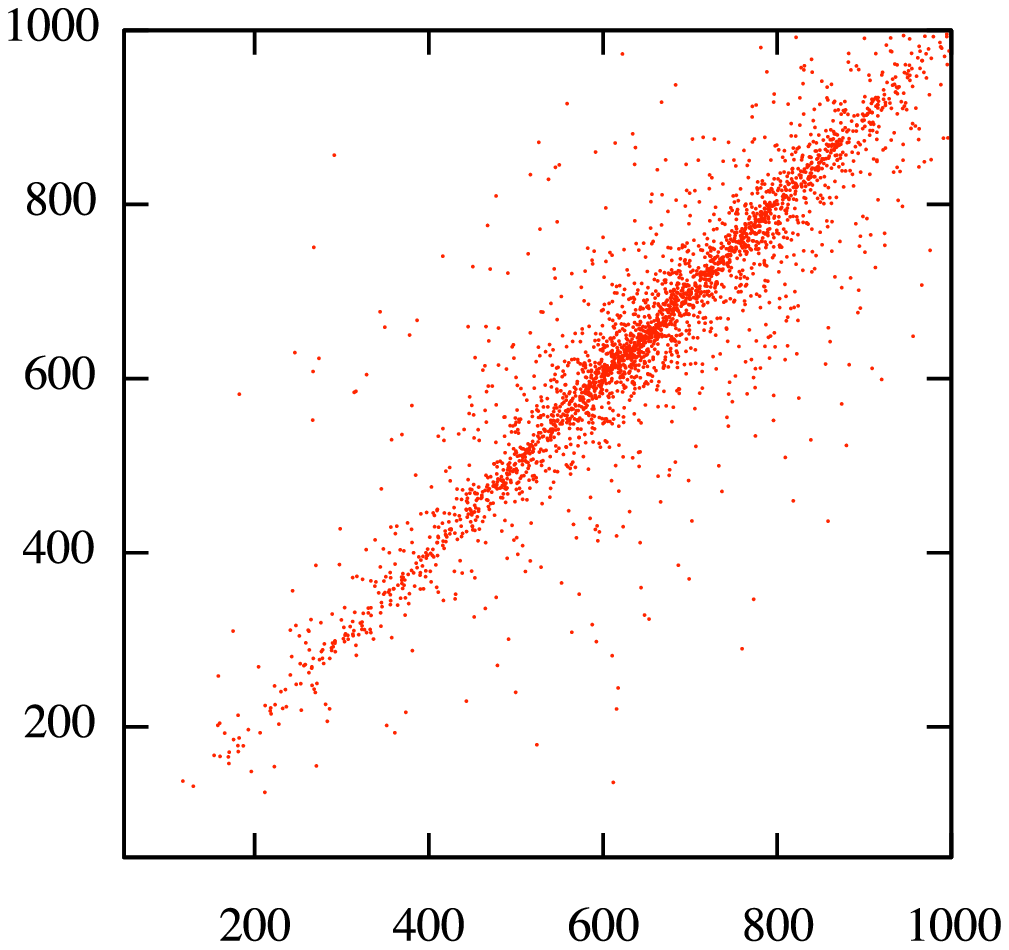}}
    \put(0.488333333333333,0.206333333333333){\rotatebox{90}{$\pt^+\;[$GeV$]$}}
    \put(0.715,0.0){$\pt^-\;[$GeV$]$}
\end{picture}
\caption{Di-stau distributions at the $14\TEV$ LHC after the preselection cuts for $\mstau=200\GEV$ (left) and
$\mstau=800\GEV$ (right). Top: Unweighted events as dots in the
$\beta^-$-$\beta^+$-plane, where $\beta^\pm$ is the velocity of
$\stau^\pm$. 
The black dashed square denotes the region selected by the velocity cut. It contains 1146 (left) and 2824 (right) events.
Bottom: Distribution in the $\pt^{+}$-$\pt^{-}$-plane of the events passing the velocity cut.}
\label{fig:distau-obs}
\end{figure}

Figure \ref{fig:sigma-mass} shows the results. The cuts on $\eta$, $\pt$,
and $\Delta\mathcal{R}_{\stau, {\rm jet}}$ have a minor impact on the
data, whereas the velocity cut lowers the signal by about one order of
magnitude but with a slightly decreasing tendency when going to very
high masses due to the increase of slower staus. The fraction of events
passing the velocity cut is $8\%$ at $\mstau=200\GEV$ and about $20\%$
at $\mstau=800\GEV$ (cf.\ figure \ref{fig:distau-obs}, top). This is
due to the fact that the parton luminosities in a
$14\TEV$ $pp$ collision begin to decrease more drastically for
center-of-mass (CM) energies above roughly $1\TEV$\@.
The top panel of figure \ref{fig:distau-obs} also shows that the velocity cut at $\beta=0.6$ 
has considerably less impact on the data than the one at $\beta=0.8$.

The partonic process considered in (\ref{eq:DYcross-section}) favors
perpendicular scattering:
$\left(\hat{u}\hat{t}-m_{\tilde{\tau}}^4\right)\D\hat{t}\propto\sin^2\theta\,\D\Omega$,
where $\theta$ is the angle between the produced staus and the beam axis
in the CM frame and $\D\Omega$ the corresponding solid angle element.
Thus, the very low $\pt$ region is suppressed. On the other
hand, the faster decrease of the parton luminosity with increasing CM
energy determines the high-$\pt$ tail. This leads to a
$\pt$-distribution that peaks roughly at $\pt\simeq\mstau$, at least
for the mass range $\mstau\lesssim 400\GEV$\@. For larger masses the behavior 
of the parton luminosity at high CM energies shifts this peak a bit downwards (see figure
\ref{fig:distau-obs}, bottom). 
Figure \ref{fig:distau-obs}, bottom also shows that the $\pt$-distributions of the two staus are clearly correlated.

%=====================
%    \                                           |
%      \                                         |
%        \                                       |
%                                      \         |
%                                        \       |
%                                          \     |
%=====================

%===========================================================================================
\section{Discovery potential and exclusion limits} \label{sec:Discovery}
%===========================================================================================

In the following we are interested in the integrated luminosity
$\IL(\mstau)$ required to discover or exclude the considered scenario characterized by the parameter $\mstau$. 
The expectation value for the number of signal events $S$ is given by
\beq
S=\sigma_{\rm S}(\mstau)\,\IL\,\epsilon\,,
\label{eq:S}
\eeq
where $\sigma_{\rm S}(\mstau)$ is the signal cross section and $\epsilon$ is the detector efficiency. 
The expected number of background events
reads
\beq
B=\sigma_{\rm B}\R\,\IL \,,
\label{eq:B}
\eeq
where $\sigma_{\rm B}$ is the background cross section and $\R$ is the background rejection factor due to the velocity discrimination.
(Conservatively, we set the detector efficiency for the background to one.)

Since we expect to obtain solutions that involve small event numbers $S$
and $B$, we consider Poisson statistics.
A $5\sigma$-discovery corresponds to a set of $S$ and $B$ that fulfills
\beq
1-\E^{-B}\sum_{n=0}^{B+S-1}\frac{B^n}{n!}\stackrel{!}=3\times10^{-7}\,,
\label{eq:disc}
\eeq
where $3\times10^{-7}$ is the one-sided $p$-value corresponding to a $5\sigma$-evidence. 

A $95\%\,$C.L. exclusion corresponds to $S$ and $B$ satisfying
\beq
1-\frac{\E^{-(B+S)}\sum_{n=0}^{N}\frac{(B+S)^n}{n!}}{\E^{-B}\sum_{n=0}^{N}\frac{B^n}{n!}}\stackrel{!}=0.95\,.
\label{eq:excl}
\eeq
In contrast to the case of discovery, the additional parameter $N$ appears. This is the maximum observed event number up to which the $95\%\,$C.L. exclusion is demanded to hold.
For $N=B$ we obtain the central value of the exclusion limit.
Repeating the analysis for $N$ at the boundaries of the $1\sigma$ and
$2\sigma$ intervals around $B$ then yields the $1\sigma$ and $2\sigma$
probability bands for the exclusion limit.

Inserting $S$ and $B$ as functions of $\IL$ and $\sigma_{\rm B}$ according to (\ref{eq:S}) and (\ref{eq:B}) turns the formulae (\ref{eq:disc}) and (\ref{eq:excl}) into implicit functions determining $\IL(\mstau)$. 

%=====================
%    \                                           |
%      \                                         |
%        \                                       |
\begin{figure}[!t]
\centering
\setlength{\unitlength}{1\textwidth}
\begin{picture}(0.78,0.55)
  \put(0.0,0.03){\includegraphics[scale=1.42, bb= 0 0 60 120, clip=true]{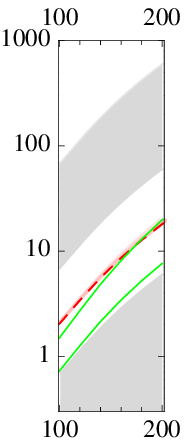}}
  \put(0.173,0.03){\includegraphics[scale=1.42, bb= 15 0 600 120, clip=true]{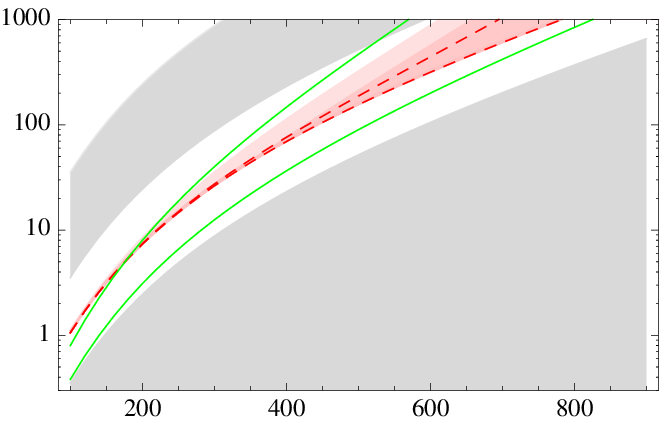}}
    \put(0.72,0.447){$\mbox{\small{${\color{green} 10^{-10}}$}}$}
    \put(0.515,0.447){$\mbox{\small{${\color{green}10^{-7}}$}}$}
    \put(0.597,0.447){$\mbox{\small{${\color{red}10^{-7}\,\,10^{-10}}$}}$}
    \put(-0.04,0.115){\rotatebox{90}{Int. Luminosity $[$fb$^{-1}]$}}
    \put(0.36,-0.006){$m_{\stau_1}\;[$GeV$]$}
    \put(0.551,0.3233){\small \color{evgray}\rotatebox{22}{-- 1 event --}}
    \put(0.251,0.2507){\small \color{evgray}\rotatebox{46}{-- 10 events}}
    \put(0.203,0.3){\small \color{evgray}\rotatebox{55.0}{-- 100 events}}  
    \put(0.087,0.081){$7\TEV$}   
    \put(0.667,0.081){$14\TEV$}
\end{picture}
\caption{Integrated luminosity at which a
$5\sigma$-discovery (green solid lines) and a $95\%\,$C.L. exclusion (red
dashed lines) of directly produced stau pairs is to be expected.
We have chosen the stau mixing angle $\thest^\text{min}$ that yields the
smallest production cross section.  The
dependence on the background rejection factor for the velocity cut is
illustrated by displaying each curve for $\R=10^{-10}$ as well as $\R=10^{-7}$.  The
dark and light red-shaded band 
around the $\R=10^{-7}$ curve displays
the $1\sigma$ and $2\sigma$ probability
band, respectively.
The results are shown for the $7\TEV$ and $14\TEV$ run of the LHC.
}
\label{fig:lum-num-mass-fix}
\end{figure}
%                                      \         |
%                                        \       |
%                                          \     |
%=====================

As signal cross section we use $\sigma_{\rm S}(\mstau)$ at
$\thest=\thest^{\rm min}$ after applying all cuts mentioned above. 
The trigger and reconstruction efficiency for single stau candidates can be conservatively
estimated to be $80\%$ \cite{Peter},
thus we choose (again conservatively) $\epsilon=0.8^2$.
The background cross section $\sigma_\text{B}$ is the di-muon cross
section after the preselection cuts.
As explained earlier, we consider the values 
$10^{-10}$ as well as $10^{-7}$ for $\R$.
Figure \ref{fig:lum-num-mass-fix} shows the luminosity $\IL(\mstau)$ 
at which a $5\sigma$-discovery can be expected (green solid lines) 
and at which all scenarios with a metastable stau can be excluded 
at $95\%\,$C.L. (red dashed lines, with red-shaded regions around
the $10^{-7}$ curve indicating the $1\sigma$ and $2\sigma$ probability bands).%
\footnote{
Since even the $2\sigma$ probability band around the 
$10^{-10}$ curve is almost degenerate with the curve itself, 
we refrain from displaying the bands for $\R=10^{-10}$.
}
The borders of the gray-shaded regions denote the luminosities that
correspond to 1, 10, and 100 events.  We see that both discovery and
exclusion are expected to occur on the basis of very few events.

Observing no events when three are expected by a hypothesis is
sufficient to exclude this hypothesis at $95\%\,$C.L.
Thus, for luminosities that lead to a sufficiently small background
$B\ll1$, the exclusion limit corresponds to the three-event line. This
is why for small $\mstau$ the exclusion limits for different $\R$ are
degenerate at the three-event line, which coincides with the
$\R=10^{-10}$ line in figure \ref{fig:lum-num-mass-fix}---the
corresponding background is sufficiently suppressed.
As it is not possible to obtain less than zero events,
the red-shaded regions do not continue beyond the three-event line.

In figure \ref{fig:lum-num-mass-fix} we also show the results for the
$7\TEV$ run of the LHC\@.  The calculation has been analogous except
that we have used a slightly smaller $K$-factor of $1.3$ for the
normalization of the stau production cross section.  We find that
as the integrated luminosity exceeds an inverse femtobarn, the LHC is
close to tightening the LEP bound of $\mstau \gtrsim 100\GEV$, which
currently remains the best limit on the direct production of long-lived sleptons.
If the $7\TEV$ run reaches $10\fb^{-1}$, one will be able to exclude
stau masses up to roughly $170\GEV$.

%-------------------------------------------------------------------------------------------
\subsection*{Optimized $\boldsymbol{\pt}$-cut}
%-------------------------------------------------------------------------------------------

Looking at the $\pt$-distribution of the staus (figure
\ref{fig:distau-obs}, bottom) and the $\pt$-cut dependence of the
di-muon cross section (figure \ref{fig:dimu}, right) we see that we can
improve the search by optimizing the $\pt$-cut according to the stau
mass hypothesis being considered.
In other words, we repeat the previous
analysis with an additional $\pt$-cut (again on both stau
candidates) that grows linearly with $\mstau$,
\beq
\pt > \pt^\text{min}=0.4\times\mstau\,.
\eeq
The stau cross section after applying this additional cut is nearly identical 
to the lowermost curve in figure \ref{fig:sigma-mass}.
We emphasize that the result is not sensitive to the choice of the
factor $0.4$. Choosing $0.5$ instead gives almost the same result. 

%=====================
%    \                                           |
%      \                                         |
%        \                                       |
\begin{figure}
\centering
\setlength{\unitlength}{1\textwidth}
\begin{picture}(0.68,0.55)
  \put(0.005,0.03){\includegraphics[scale=1.42, bb= 0 0 600 120, clip=true]{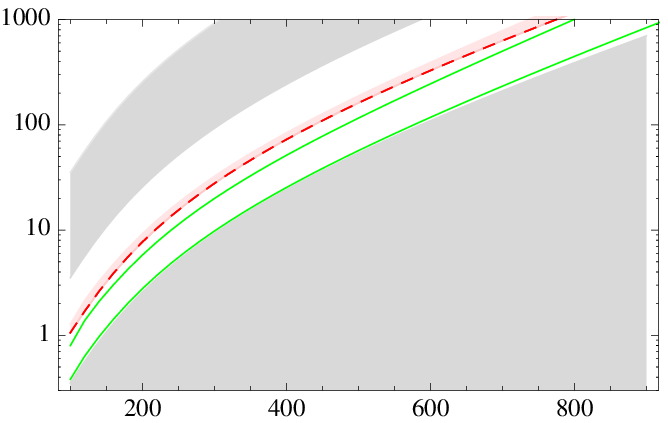}} 
    \put(0.63,0.447){$\mbox{\small{${\color{green} 10^{-10}}$}}$}
    \put(0.56,0.447){$\mbox{\small{${\color{green}10^{-7}}$}}$}
    \put(-0.038,0.116){\rotatebox{90}{Int. Luminosity $[$fb$^{-1}]$}}
    \put(0.303,-0.006){$m_{\stau_1}\;[$GeV$]$}
    \put(0.43,0.3233){\small \color{evgray}\rotatebox{22}{-- 1 event --}}
    \put(0.13,0.2507){\small \color{evgray}\rotatebox{46}{-- 10 events}}
    \put(0.086,0.3){\small \color{evgray}\rotatebox{55.0}{-- 100 events}}  
    \put(0.546,0.081){$14\TEV$}
\end{picture}
\caption{Same as figure \ref{fig:lum-num-mass-fix}, but with an
additional cut $\pt>0.4\,\mstau$.}
\label{fig:lum-num-mass-run}
\end{figure}
%                                      \         |
%                                        \       |
%                                          \     |
%=====================

The dependence of the background cross section on this optimized
$\pt$-cut is shown in the right panel of figure \ref{fig:dimu}. Apart
from lower statistics, in the high-$\pt$ region we expect a larger
theoretical uncertainty. Therefore we perform a conservative fit to the
data, such that the fitted curve lies completely above the simulated
curve in the region $\pt>50\GEV$.

After this change, discovery is expected to take place for one to three events in the whole
considered region, as shown in figure~\ref{fig:lum-num-mass-run}.
For instance, a scenario with a $300\GEV$ stau NLSP
is expected to be discovered at about $10\fb^{-1}$ 
(and even for $\R=10^{-7}$ at about $20\fb^{-1}$)
through direct production alone. On the other hand, if it is not chosen by nature
the same scenario can be excluded at $95\%\,$C.L. with roughly $30\fb^{-1}$.
In the very long term, nearly the whole considered mass range is
accessible at the LHC, for example, masses up to $600\GEV$ (exclusion) 
and about $750\GEV$ (discovery) for $300\fb^{-1}$. 
Note that the probability bands for the $\R=10^{-10}$ exclusion curve 
are degenerate with the three-event line.
In figure \ref{fig:lum-num-mass-run} only the 
$2\sigma$ probability band for the $10^{-7}$ curve is visible.

Besides, the dependence of the LHC potential on the background rejection
factor $\R$ can be studied. The large change of $\R$ by three orders of
magnitude causes only a small change in the luminosity
required for a discovery by a factor of about $2$.
Note that the discovery curves are the expected discovery reach. Since
for $\R=10^{-10}$ the sufficient event number for discovery is one or
two, the statistical fluctuation is of $\mathcal{O}(1)$, too.
So a discovery can easily take place at half or double the integrated
luminosity shown in the figures.%
\footnote{For this reason we refrain from quantizing the required event
numbers to integers.}
Thus, the difference between the $\R=10^{-7}$ and $\R=10^{-10}$
discovery curve is of the same order as the expected statistical
fluctuation. For this reason we assess the discovery curves to be
relatively insensitive to the change in $\R$ in the considered range.
Moreover, the exclusion limits for $\R=10^{-7}$ and $10^{-10}$ become degenerate at the three-event line.
Thus, the exclusion potential is not sensitive to $\R$ at all.

This also sheds some light on the impact of the uncertainties in the background cross section. 
The variation of $\R$ over three orders of magnitude reveals 
how little impact an uncertainty of tens or even hundreds of percent
in the background cross section would have. This justifies the 
somewhat arbitrarily chosen $K$-factor for the muon background in section 
\ref{sec:background} and shows that setting the detector efficiency 
for the background to one was not very conservative. 
Similarly, it shows that our result does not rely on the exact shape of the high 
$\pt$ tail of the distribution in the right panel of figure \ref{fig:dimu}. 

To realize this optimized $\pt$-cut in an experimental
analysis, one could choose the value of $\pt^\text{min}$ corresponding
to the stau mass that---for example, according to figure
\ref{fig:lum-num-mass-run}---is within reach
at the luminosity of the analyzed dataset.

%===========================================================================================
\section{Direct production versus other channels}\label{sec:production}
%===========================================================================================

The cross section for the production of superparticles heavier than the
stau can be larger than the DY cross section considered so far.  Such
sparticles promptly decay into the NLSP, emitting Standard Model (SM)
particles which provide an additional signature with potentially higher
significance than the detection of the metastable stau. However, the
SM particle radiation from cascade decays depends strongly on the
sparticle mass spectrum and has to be distinguished from a higher SM
background.
Hence, we assume that the easiest way to find SUSY in the considered
scenario is the direct detection of the NLSP, independently of its
production channel.  Then we can estimate whether staus from direct
production or those from decays of heavier sparticles will be the
dominant contribution to the discovery of a gravitino-stau scenario.
This enables us to decide for which SUSY spectra our previous
calculations for the direct DY production yield a good approximation for
the potential of the LHC and for which spectra they are overly
conservative.

We classify three sources of the production of SUSY particles,
\begin{itemize}
\item
production of sleptons, including the direct production of staus,
\item
production of neutralinos and charginos, and
\item
production of colored sparticles.
\end{itemize}
Let us look at the leading contributions of each class. 
For simplicity, in each case we consider exemplary production rates
of a single sparticle species and do not sum, e.g., over the generations 
of sfermions 
(which would require an assumption on the relation between their masses). 
It is easy to estimate the production rate in the case of degeneracies 
by multiplying the results by the appropriate factor.
For this
consideration we computed the cross section via 
\textsc{MadGraph/MadEvent 5} \cite{Alwall:2011uj}
at lowest order and cross-checked whether NLO computations
\cite{PhysRevD.57.5871,1999PhRvL..83.3780B,1997NuPhB.492...51B} lead to
roughly the same conclusions.

At lowest order in the electroweak coupling, $\mathcal{O}(\alpha^2)$,
direct production of sleptons is only possible via neutral current and
charged current DY\@.
According to figure \ref{fig:theta-stau}, right- and left-handed
slepton pair production differs by a factor of about $2$ in the cross
section. On the other hand, such an increase of the cross section
is compensated already by a rather small increase of the slepton mass by a factor
of less than about $1.3$. 
Accordingly, the contribution from another neutral current DY produced slepton 
$\SY{l}_{\text{L}}$ decaying into the NLSP will only be noticeable if 
$\stau_1\simeq\stau_{\text{R}}$ and for a very small gap between 
$\mstau$ and $m_{\SY{l}_{\text{L}}}$.
Production of $\SY{l}_{\text{L}}\,\SY{\nu}_{l}$ via $W^{\pm}$ is in principle
enhanced relative to $Z,\gamma\to\SY{l}^{+}_{\text{L}}\,\SY{l}^{-}_{\text{L}}$ by a
factor of $2$ to $3$. Since only $\SY{l}_{\text{L}}$ couples to $W^{\pm}$, direct
production of $\stau_1$ via $W^{\pm}$ is suppressed by $\sin^2\thest$
and thus unlikely for $\stau_1\simeq\stau_{\text{R}}$. By contrast, production of
$\SY{l}^{+}_{\text{L}}\,\SY{\nu}_{l}$ via $W^{+}$ can be of the same order as 
$Z,\gamma\to\stau^{+}_{\text{R}}\,\stau^{-}_{\text{R}}$ if $m_{\SY{l}_{\text{L}}}\simeq m_{\SY{\nu_{l}}}\lesssim1.6\,\mstau$. 
(The production via $W^+$ is slightly enhanced against the one via $W^-$ due to the charge asymmetry in the initial state.)
However, in any case, the production of another slepton pair 
$\SY{l}^{\pm}\,\SY{\nu}_{l}$ or $\SY{l}^{+}\SY{l}^{-}$ 
decaying into a pair of $\stau_1$ does not have the potential to exceed the direct DY production drastically.

Production of neutralinos and charginos is accessible at
$\mathcal{O}(\alpha^2)$ either via neutral and charged current DY or via
$t$- and $u$-channel squark exchange.
DY production of neutralinos only occurs in the case of a noticeable
contribution of higgsinos and even then the cross sections are quite
small. In contrast, $Z,\gamma\to\SY{W}^{+}\,\SY{W}^{-}$ has quite a large cross section. The same is true for  
$W^{\pm}\to\SY{W}^{\pm}\,\SY{W}^{0}$. For
$m_{\SY{W}^{\pm}}\lesssim3\,m_{\stau_{\text{R}}}$ the DY production of
$\SY{W}^{+}\,\SY{W}^{-}$ can exceed
$Z,\gamma\to\stau^{+}_{\text{R}}\,\stau^{-}_{\text{R}}$. If the charginos are more
higgsino-like the cross sections become lower,
$m_{\SY{H}^{\pm}}\lesssim2\,m_{\stau_{\text{R}}}$ is required for a competing
or dominant higgsino production. The production via squarks in the $t$- and
$u$-channel introduces an additional dependence on the squark masses.
However, the cross sections are roughly the same as the corresponding
production via DY only if $m_{\SY{q}}\simeq m_{\SY{W}}$. Thus, for larger
squark masses the $t$- and $u$-channel squark contributions become
subleading. According to these considerations, 
one should keep in mind the chargino production via DY, which can become an important
channel, especially in the case of a light wino.

Let us finally consider the production of colored sparticles. Since the
LHC is a proton-proton collider, at leading order
$\mathcal{O}(\alpha_{s}^{2})$ squark-pair, gluino-pair and squark-gluino
production each allow a variety of diagrams that include the
initial states $gg$, $qg$ and $qq$, which are the dominant hadronic
channels at the LHC for low, middle and high CM energies, respectively.
Setting all masses equal to $1\TEV$, the production of the colored 
sparticles and direct stau-pair production follow roughly the ratio 
\beq
\sigma_{\SY{q}\SY{g}}:\sigma_{\SY{g}\SY{g}}:\sigma_{\,\overline{\SY{q}\,}\!\SY{q}}:\sigma_{\stau^{+}_{\text{R}}\stau^{-}_{\text{R}}}\,
\simeq\;{1}\,:\,\frac{1}{16}\,:\,\frac{1}{33}\,:\,\frac{1}{1700}\,,
\label{eq:ratio}
\eeq
where we chose $\SY{q}=\SY{u}$ (either right- or left-handed), which is the leading contribution. 
Thus, spectra with dominant direct DY production of staus are required to have
quite a large mass gap between the NLSP and the colored sparticles 
to compensate this ratio.
Large mass gaps are typical for gauge
\cite{Dine:1981gu,AlvarezGaume:1981wy} or gaugino mediated
\cite{Kaplan:1999ac,Chacko:1999mi} SUSY breaking, and no-scale models
\cite{Ellis:1984bm}, for example.
Furthermore, (\ref{eq:ratio}) shows that gluino-squark production is most likely to be the dominant
contribution to the production of colored sparticles.%
\footnote{As long as the masses of the squarks are not vastly different,
this conclusion remains to hold if we sum over the squark flavors 
and vary the overall mass scale within reasonable boundaries.}

Let us close these considerations with two examples. 
In a scenario with 
\bea 
m_{\stau_{\text{R}}}=200\GEV\,,\; m_{\SY{W}^{\pm}}=600\GEV\,,\;m_{\SY{u}}=1.4\TEV\,,\;m_{\SY{g}}=1.8\TEV
\eea
or, for a larger overall mass scale,
\bea 
m_{\stau_{\text{R}}}=800\GEV\,,\; m_{\SY{W}^{\pm}}=1.7\TEV\,,\;m_{\SY{u}}=2.6\TEV\,,\;m_{\SY{g}}=3.2\TEV
\,,
\eea
the cross sections for
$\stau_{\text{R}}^{+}\stau_{\text{R}}^{-}$, $\SY{W}^{+}\SY{W}^{-}$ and $\SY{u}\SY{g}$
production are all roughly of the same size. 

However, in the considered case of
a rather large mass gap to the colored sparticles,
the staus from cascade decays have significantly
higher velocities than directly produced ones (due to the large phase
space) \cite{Rajaraman:2006mr,ourCascadePaper}. 
Consequently, staus from cascade decays are more likely to be
rejected by the velocity cut needed for the discrimination against muons.
Thus, already in the exemplary equal-production-rate scenarios from
above we expect direct DY production to be the dominant contribution to
detectable staus.

%===========================================================================================
\section{Conclusions}
%===========================================================================================

Metastable charged supersymmetric particles lead to prominent signatures
in the detectors at the LHC\@. We have shown that these signatures
enable a very efficient background rejection.  
As a consequence, despite
its relatively small cross section direct Drell-Yan production of
metastable charged sleptons has an interesting potential for discovering
or excluding their existence at the LHC for a wide range of masses.
Above all, it provides a robust lower limit on the LHC potential for
scenarios with a metastable charged slepton that depends only on the
slepton mass.
For instance, the $7\TEV$ run will improve the LEP limit in the near
future and could exclude slepton masses up to roughly $160\GEV$ with an
integrated luminosity of $10\fb^{-1}$.

Particularly for the heavy mass range, we have proposed an additional
cut depending on the slepton mass which may further reduce the
background.  At the $14\TEV$ LHC, this would
allow to discover a $300\GEV$ stau, for example, at about $10\fb^{-1}$
via direct production.
With a very large
luminosity of $300\fb^{-1}$, masses up to roughly $600\GEV$ could be
excluded, and even heavier staus could be discovered.
As mentioned, this mass region can be regarded as interesting
from a cosmological point of view due to the constraints from catalyzed 
big bang nucleosynthesis.

In the spirit of a model-independent approach, we have assumed the stau
mixing angle yielding the minimal production cross section, which is
slightly below the cross section for a pure $\stau_\text{R}$.  The
limits on $\stau_\text{L}$ are correspondingly tighter due to its larger
production cross section.  Concerning experimental issues, we chose to
be conservative as well.  The LHC potential may improve, for example,
with a better control over the distinction between charged sleptons and
muons or a better detector efficiency than assumed here.

By considering channels other than direct Drell-Yan production, a
larger, albeit more model-dependent, discovery potential and tighter
exclusion
limits can be achieved. We have discussed briefly the production
processes that are most likely to be the dominant ones if the mass gap
between the metastable charged slepton and the heavier superparticles is
not large enough to guarantee dominant Drell-Yan production. At the LHC,
for many mass spectra this would be wino production or associated
squark-gluino production.

\subsection*{Acknowledgements}
We would like to thank Christian Autermann, Patrick Huber, Kolja
Kaschube, Boris Panes, Christian Sander, Peter Schleper and Hartmut
Stadie for very helpful discussions.
We also thank CINVESTAV in Mexico City for hospitality during stages
of this work.
This work was supported by the German Science Foundation (DFG) via the
Junior Research Group ``SUSY Phenomenology'' within the Collaborative
Research Centre 676 ``Particles, Strings and the Early Universe''.

\phantomsection % Ensures that a PDF bookmark is set here
\addcontentsline{toc}{chapter}{References}
\bibliographystyle{../utphys}
\bibliography{../staus}

\providecommand{\href}[2]{#2}\begingroup\raggedright\begin{thebibliography}{10}

\bibitem{Fayet:1981sq}
P.~Fayet, ``Experimental consequences of supersymmetry'', in {\em Proceedings
  of the 16th Rencontre de Moriond}, J.~{Tran Thanh Van}, ed., vol.~1,
  pp.~347--367.
\newblock Editions Frontieres,
1981.
\newblock
%%CITATION = C81-03-15-25;%%.

\bibitem{Pagels:1981ke}
H.~Pagels and J.~R. Primack, ``{Supersymmetry, Cosmology and New TeV
  Physics}'',
\href{http://dx.doi.org/10.1103/PhysRevLett.48.223}{{\em Phys. Rev. Lett.} {\bf
  48} (1982)  223}.
%%CITATION = PRLTA,48,223;%%.

\bibitem{Ellis:1984er}
J.~R. Ellis, D.~V. Nanopoulos, and S.~Sarkar, ``{The Cosmology of Decaying
  Gravitinos}'',
\href{http://dx.doi.org/10.1016/0550-3213(85)90306-2}{{\em Nucl. Phys.} {\bf
  B259} (1985)  175}.
%%CITATION = NUPHA,B259,175;%%.

\bibitem{Bolz:1998ek}
M.~Bolz, W.~Buchm{\"u}ller, and M.~Pl{\"u}macher, ``{Baryon asymmetry and dark
  matter}'', \href{http://dx.doi.org/10.1016/S0370-2693(98)01342-2}{{\em Phys.
  Lett.} {\bf B443} (1998)  209--213},
\href{http://arxiv.org/abs/hep-ph/9809381}{{\tt arXiv:hep-ph/9809381}}.
%%CITATION = HEP-PH/9809381;%%.

\bibitem{Brandenburg:2005he}
A.~Brandenburg, L.~Covi, K.~Hamaguchi, L.~Roszkowski, and F.~D. Steffen,
  ``Signatures of axinos and gravitinos at colliders'',
  \href{http://dx.doi.org/10.1016/j.physletb.2005.04.072}{{\em Phys. Lett.}
  {\bf B617} (2005)  99--111},
\href{http://arxiv.org/abs/hep-ph/0501287}{{\tt arXiv:hep-ph/0501287}}.
%%CITATION = HEP-PH 0501287;%%.

\bibitem{Jittoh:2005pq}
T.~Jittoh, J.~Sato, T.~Shimomura, and M.~Yamanaka, ``Long life stau in the
  minimal supersymmetric standard model'',
  \href{http://dx.doi.org/10.1103/PhysRevD.73.055009}{{\em Phys. Rev.} {\bf
  D73} (2006)  055009},
\href{http://arxiv.org/abs/hep-ph/0512197}{{\tt arXiv:hep-ph/0512197}}.
%%CITATION = HEP-PH 0512197;%%.

\bibitem{Buchmuller:2004rq}
W.~Buchm{\"u}ller, K.~Hamaguchi, M.~Ratz, and T.~Yanagida, ``Supergravity at
  colliders'', \href{http://dx.doi.org/10.1016/j.physletb.2004.03.016}{{\em
  Phys. Lett.} {\bf B588} (2004)  90--98},
\href{http://arxiv.org/abs/hep-ph/0402179}{{\tt arXiv:hep-ph/0402179}}.
%%CITATION = HEP-PH 0402179;%%.

\bibitem{Pospelov:2006sc}
M.~Pospelov, ``Particle physics catalysis of thermal {B}ig {B}ang
  {N}ucleosynthesis'',
  \href{http://dx.doi.org/10.1103/PhysRevLett.98.231301}{{\em Phys. Rev. Lett.}
  {\bf 98} (2007)  231301},
\href{http://arxiv.org/abs/hep-ph/0605215}{{\tt arXiv:hep-ph/0605215}}.
%%CITATION = HEP-PH/0605215;%%.

\bibitem{Pradler:2007is}
J.~Pradler and F.~D. Steffen, ``Implications of catalyzed {BBN} in the {CMSSM}
  with gravitino dark matter'',
  \href{http://dx.doi.org/10.1016/j.physletb.2008.07.050}{{\em Phys. Lett.}
  {\bf B666} (2008)  181--184},
\href{http://arxiv.org/abs/0710.2213}{{\tt arXiv:0710.2213 [hep-ph]}}.
%%CITATION = 0710.2213;%%.

\bibitem{Kersten:2007ab}
J.~Kersten and K.~Schmidt-Hoberg, ``The gravitino-stau scenario after catalyzed
  {BBN}'', \href{http://dx.doi.org/10.1088/1475-7516/2008/01/011}{{\em JCAP}
  {\bf 0801} (2008)  011},
\href{http://arxiv.org/abs/0710.4528}{{\tt arXiv:0710.4528 [hep-ph]}}.
%%CITATION = 0710.4528;%%.

\bibitem{Nisati:1997gb}
A.~Nisati, S.~Petrarca, and G.~Salvini, ``{On the possible detection of massive
  stable exotic particles at the LHC}'',
  \href{http://dx.doi.org/10.1142/S0217732397002260}{{\em Mod. Phys. Lett.}
  {\bf A12} (1997)  2213--2222},
  \href{http://arxiv.org/abs/hep-ph/9707376}{{\tt arXiv:hep-ph/9707376}}.

\bibitem{Hinchliffe:1998ys}
I.~Hinchliffe and F.~Paige, ``{Measurements in gauge mediated SUSY breaking
  models at CERN LHC}'',
  \href{http://dx.doi.org/10.1103/PhysRevD.60.095002}{{\em Phys. Rev.} {\bf
  D60} (1999)  095002}, \href{http://arxiv.org/abs/hep-ph/9812233}{{\tt
  arXiv:hep-ph/9812233}}.

\bibitem{Ambrosanio:2000ik}
S.~Ambrosanio, B.~Mele, S.~Petrarca, G.~Polesello, and A.~Rimoldi, ``{Measuring
  the SUSY breaking scale at the LHC in the slepton NLSP scenario of GMSB
  models}'', \href{http://dx.doi.org/10.1088/1126-6708/2001/01/014}{{\em JHEP}
  {\bf 01} (2001)  014},
\href{http://arxiv.org/abs/hep-ph/0010081}{{\tt arXiv:hep-ph/0010081}}.
%%CITATION = HEP-PH/0010081;%%.

\bibitem{Ambrosanio:2000zu}
S.~Ambrosanio, B.~Mele, A.~Nisati, S.~Petrarca, G.~Salvini, G.~Polesello, and
  A.~Rimoldi, ``{SUSY Long-Lived Massive Particles: Detection and Physics at
  the LHC}'', \href{http://dx.doi.org/10.1007/BF02904518}{{\em Rendiconti
  Lincei} {\bf 12} (2001)  5--18},
\href{http://arxiv.org/abs/hep-ph/0012192}{{\tt arXiv:hep-ph/0012192}}.
%%CITATION = HEP-PH/0012192;%%.

\bibitem{Sjolin:2002fx}
J.~Sj{\"o}lin, ``{A simulation of gauge mediated supersymmetry breaking with a
  supersymmetric tau as the next-to-lightest supersymmetric particle in the
  ATLAS detector at the large hadron collider}'',
  \href{http://dx.doi.org/10.1007/s1010502c0011}{{\em EPJ direct} {\bf 4}
  (2002)  1--23}.

\bibitem{Feng:2004mt}
J.~L. Feng, S.~Su, and F.~Takayama, ``{Supergravity with a gravitino LSP}'',
  \href{http://dx.doi.org/10.1103/PhysRevD.70.075019}{{\em Phys. Rev.} {\bf
  D70} (2004)  075019},
\href{http://arxiv.org/abs/hep-ph/0404231}{{\tt arXiv:hep-ph/0404231}}.
%%CITATION = HEP-PH/0404231;%%.

\bibitem{DeRoeck:2005bw}
A.~{De Roeck}, J.~Ellis, F.~Gianotti, F.~Moortgat, K.~A. Olive, and L.~Pape,
  ``{Supersymmetric benchmarks with non-universal scalar masses or gravitino
  dark matter}'', \href{http://dx.doi.org/10.1140/epjc/s10052-006-0182-6}{{\em
  Eur. Phys. J.} {\bf C49} (2007)  1041--1066},
\href{http://arxiv.org/abs/hep-ph/0508198}{{\tt arXiv:hep-ph/0508198}}.
%%CITATION = HEP-PH/0508198;%%.

\bibitem{Ellis:2006vu}
J.~R. Ellis, A.~R. Raklev, and O.~K. {\O}ye, ``{Gravitino dark matter scenarios
  with massive metastable charged sparticles at the LHC}'',
  \href{http://dx.doi.org/10.1088/1126-6708/2006/10/061}{{\em JHEP} {\bf 10}
  (2006)  061},
\href{http://arxiv.org/abs/hep-ph/0607261}{{\tt arXiv:hep-ph/0607261}}.
%%CITATION = HEP-PH/0607261;%%.

\bibitem{Feng:2009yq}
J.~L. Feng, S.~T. French, C.~G. Lester, Y.~Nir, and Y.~Shadmi, ``{The Shifted
  Peak: Resolving Nearly Degenerate Particles at the LHC}'',
  \href{http://dx.doi.org/10.1103/PhysRevD.80.114004}{{\em Phys. Rev.} {\bf
  D80} (2009)  114004},
\href{http://arxiv.org/abs/0906.4215}{{\tt arXiv:0906.4215 [hep-ph]}}.
%%CITATION = 0906.4215;%%.

\bibitem{Feng:2009bd}
J.~L. Feng, S.~T. French, I.~Galon, C.~G. Lester, Y.~Nir, Y.~Shadmi,
  D.~Sanford, and F.~Yu, ``{Measuring slepton masses and mixings at the LHC}'',
  \href{http://dx.doi.org/10.1007/JHEP01(2010)047}{{\em JHEP} {\bf 01} (2010)
  047},
\href{http://arxiv.org/abs/0910.1618}{{\tt arXiv:0910.1618 [hep-ph]}}.
%%CITATION = 0910.1618;%%.

\bibitem{Chen:2009gu}
J.~Chen and T.~Adams, ``{Searching for high speed long-lived charged massive
  particles at the LHC}'',
  \href{http://dx.doi.org/10.1140/epjc/s10052-010-1283-9}{{\em Eur. Phys. J.}
  {\bf C67} (2010)  335--342},
\href{http://arxiv.org/abs/0909.3157}{{\tt arXiv:0909.3157 [hep-ph]}}.
%%CITATION = 0909.3157;%%.

\bibitem{Ito:2009xy}
T.~Ito, R.~Kitano, and T.~Moroi, ``{Measurement of the superparticle mass
  spectrum in the long-lived stau scenario at the LHC}'',
  \href{http://dx.doi.org/10.1007/JHEP04(2010)017}{{\em JHEP} {\bf 04} (2010)
  017},
\href{http://arxiv.org/abs/0910.5853}{{\tt arXiv:0910.5853 [hep-ph]}}.
%%CITATION = 0910.5853;%%.

\bibitem{Heckman:2010xz}
J.~J. Heckman, J.~Shao, and C.~Vafa, ``{F-theory and the LHC: stau search}'',
  \href{http://dx.doi.org/10.1007/JHEP09(2010)020}{{\em JHEP} {\bf 09} (2010)
  020},
\href{http://arxiv.org/abs/1001.4084}{{\tt arXiv:1001.4084 [hep-ph]}}.
%%CITATION = 1001.4084;%%.

\bibitem{Kitano:2010tt}
R.~Kitano and M.~Nakamura, ``{Tau polarization measurements at the LHC in
  supersymmetric models with a long-lived stau}'',
  \href{http://dx.doi.org/10.1103/PhysRevD.82.035007}{{\em Phys. Rev.} {\bf
  D82} (2010)  035007},
\href{http://arxiv.org/abs/1006.2904}{{\tt arXiv:1006.2904 [hep-ph]}}.
%%CITATION = 1006.2904;%%.

\bibitem{Endo:2010ya}
M.~Endo, K.~Hamaguchi, and K.~Nakaji, ``{Probing High Reheating Temperature
  Scenarios at the LHC with Long-Lived Staus}'',
  \href{http://dx.doi.org/10.1007/JHEP11(2010)004}{{\em JHEP} {\bf 11} (2010)
  004},
\href{http://arxiv.org/abs/1008.2307}{{\tt arXiv:1008.2307 [hep-ph]}}.
%%CITATION = 1008.2307;%%.

\bibitem{Endo:2011uw}
M.~Endo, K.~Hamaguchi, and K.~Nakaji, ``{LHC signature with long-lived stau in
  high reheating temperature scenario}'',
\href{http://arxiv.org/abs/1105.3823}{{\tt arXiv:1105.3823 [hep-ph]}}.
%%CITATION = 1105.3823;%%.

\bibitem{Fairbairn:2006gg}
M.~Fairbairn, A.~C. Kraan, D.~A. Milstead, T.~Sj{\"o}strand, P.~Skands, and
  T.~Sloan, ``{Stable massive particles at colliders}'',
  \href{http://dx.doi.org/10.1016/j.physrep.2006.10.002}{{\em Phys. Rept.} {\bf
  438} (2007)  1--63},
\href{http://arxiv.org/abs/hep-ph/0611040}{{\tt arXiv:hep-ph/0611040}}.
%%CITATION = HEP-PH/0611040;%%.

\bibitem{Rajaraman:2006mr}
A.~Rajaraman and B.~T. Smith, ``{Discovering SUSY with $m_0^2 < 0$ in the first
  CERN LHC physics run}'',
  \href{http://dx.doi.org/10.1103/PhysRevD.75.115015}{{\em Phys. Rev.} {\bf
  D75} (2007)  115015},
\href{http://arxiv.org/abs/hep-ph/0612235}{{\tt arXiv:hep-ph/0612235}}.
%%CITATION = HEP-PH/0612235;%%.

\bibitem{Rajaraman:2007ae}
A.~Rajaraman and B.~T. Smith, ``{Determining Spins of Metastable Sleptons at
  the Large Hadron Collider}'',
  \href{http://dx.doi.org/10.1103/PhysRevD.76.115004}{{\em Phys. Rev.} {\bf
  D76} (2007)  115004},
\href{http://arxiv.org/abs/0708.3100}{{\tt arXiv:0708.3100 [hep-ph]}}.
%%CITATION = 0708.3100;%%.

\bibitem{Raklev:2009mg}
A.~R. Raklev, ``{Massive Metastable Charged (S)Particles at the LHC}'',
  \href{http://dx.doi.org/10.1142/S0217732309031648}{{\em Mod. Phys. Lett.}
  {\bf A24} (2009)  1955--1969},
\href{http://arxiv.org/abs/0908.0315}{{\tt arXiv:0908.0315 [hep-ph]}}.
%%CITATION = 0908.0315;%%.

\bibitem{alwall-2007}
J.~Alwall, P.~Demin, S.~{de Visscher}, R.~Frederix, M.~Herquet, F.~Maltoni,
  T.~Plehn, D.~L. Rainwater, and T.~Stelzer, ``{MadGraph/MadEvent v4: The New
  Web Generation}'',
  \href{http://dx.doi.org/10.1088/1126-6708/2007/09/028}{{\em JHEP} {\bf 0709}
  (2007)  028}, \href{http://arxiv.org/abs/0706.2334}{{\tt arXiv:0706.2334
  [hep-ph]}}.

\bibitem{sjostrand-2006-0605}
T.~Sjostrand, S.~Mrenna, and P.~Skands, ``Pythia 6.4 physics and manual'',
  \href{http://dx.doi.org/10.1088/1126-6708/2006/05/026}{{\em JHEP} {\bf 0605}
  (2006)  026}, \href{http://arxiv.org/abs/hep-ph/0603175}{{\tt
  arXiv:hep-ph/0603175}}.

\bibitem{alwall-2008-53}
J.~Alwall, S.~Hoeche, F.~Krauss, N.~Lavesson, L.~Lonnblad, F.~Maltoni, M.~L.
  Mangano, M.~Moretti, C.~G. Papadopoulos, F.~Piccinini, S.~Schumann,
  M.~Treccani, J.~Winter, and M.~Worek, ``Comparative study of various
  algorithms for the merging of parton showers and matrix elements in hadronic
  collisions'', \href{http://dx.doi.org/10.1140/epjc/s10052-007-0490-5}{{\em
  Eur. Phys. J.} {\bf C53} (2008)  473--500},
  \href{http://arxiv.org/abs/0706.2569}{{\tt arXiv:0706.2569 [hep-ph]}}.

\bibitem{Pumplin:2002vw}
J.~Pumplin, D.~R. Stump, J.~Huston, H.~L. Lai, P.~Nadolsky, and W.~K. Tung,
  ``{New generation of parton distributions with uncertainties from global QCD
  analysis}'', \href{http://dx.doi.org/10.1088/1126-6708/2002/07/012}{{\em
  JHEP} {\bf 0207} (2002)  012},
  \href{http://arxiv.org/abs/hep-ph/0201195}{{\tt arXiv:hep-ph/0201195}}.

\bibitem{Albajar:687291}
C.~Albajar and G.~Wrochna, ``Isolated muon trigger'', Tech. Rep.
  CMS-NOTE-2000-067, CERN, Geneva, Sep, 2000.
\newblock \url{http://cdsweb.cern.ch/record/687291/files/note00_067.pdf}.

\bibitem{Amapane:687467}
N.~Amapane, M.~Fierro, and M.~Konecki, ``High level trigger algorithms for muon
  isolation'', Tech. Rep. CMS-NOTE-2002-040, CERN, Geneva, Nov, 2002.
\newblock \url{http://cdsweb.cern.ch/record/876167/files/0508194.pdf}.

\bibitem{PoS2008LHC012}
K.~Nawrocki, ``Search for heavy stable charged particles in {CMS}'', in {\em
  Proceedings of Physics at LHC 2008}, no.~PoS(2008LHC)012.
\newblock 2008.
\newblock
  \url{http://pos.sissa.it/archive/conferences/055/012/2008LHC_012.pdf}.

\bibitem{CMS-PAS-EXO-08-003}
{{CMS} Collaboration}, ``Search for heavy stable charged particles with 100
  pb$^{-1}$ and 1 fb$^{-1}$ in the {CMS} experiment'', Tech. Rep.
  CMS-PAS-EXO-08-003, CERN, Geneva, Feb, 2009.
\newblock \url{http://cdsweb.cern.ch/record/1152570/files/EXO-08-003-pas.pdf}.

\bibitem{PhysRevD.57.5871}
H.~Baer, B.~Harris, and M.~H. Reno, ``Next-to-leading order slepton pair
  production at hadron colliders'',
  \href{http://dx.doi.org/10.1103/PhysRevD.57.5871}{{\em Phys. Rev.} {\bf D57}
  (1998)  5871--5874}, \href{http://arxiv.org/abs/hep-ph/9712315}{{\tt
  arXiv:hep-ph/9712315}}.

\bibitem{1999PhRvL..83.3780B}
W.~{Beenakker}, M.~{Klasen}, M.~{Kr{\"a}mer}, T.~{Plehn}, M.~{Spira}, and P.~M.
  {Zerwas}, ``{Production of Charginos, Neutralinos, and Sleptons at Hadron
  Colliders}'', \href{http://dx.doi.org/10.1103/PhysRevLett.83.3780}{{\em Phys.
  Rev. Lett.} {\bf 83} (1999)  3780--3783},
  \href{http://arxiv.org/abs/hep-ph/9906298}{{\tt arXiv:hep-ph/9906298}}.
  Erratum-ibid.\ {\bf 100} (2008) 029901.

\bibitem{Bozzi:2007qr}
G.~Bozzi, B.~Fuks, and M.~Klasen, ``{Threshold Resummation for Slepton-Pair
  Production at Hadron Colliders}'',
  \href{http://dx.doi.org/10.1016/j.nuclphysb.2007.03.052}{{\em Nucl. Phys.}
  {\bf B777} (2007)  157--181}, \href{http://arxiv.org/abs/hep-ph/0701202}{{\tt
  arXiv:hep-ph/0701202}}.

\bibitem{Bozzi:2007tea}
G.~Bozzi, B.~Fuks, and M.~Klasen, ``{Joint resummation for slepton pair
  production at hadron colliders}'',
  \href{http://dx.doi.org/10.1016/j.nuclphysb.2007.10.021}{{\em Nucl. Phys.}
  {\bf B794} (2008)  46--60}, \href{http://arxiv.org/abs/0709.3057}{{\tt
  arXiv:0709.3057 [hep-ph]}}.

\bibitem{Peter}
K.~Kaschube and P.~Schleper, {\em {Personal communication}}, 2011.

\bibitem{Alwall:2011uj}
J.~Alwall, M.~Herquet, F.~Maltoni, O.~Mattelaer, and T.~Stelzer, ``{MadGraph 5:
  Going Beyond}'', \href{http://dx.doi.org/10.1007/JHEP06(2011)128}{{\em JHEP}
  {\bf 1106} (2011)  128}, \href{http://arxiv.org/abs/1106.0522}{{\tt
  arXiv:1106.0522 [hep-ph]}}.

\bibitem{1997NuPhB.492...51B}
W.~Beenakker, R.~H{\"o}pker, M.~Spira, and P.~Zerwas, ``{Squark and gluino
  production at hadron colliders}'',
  \href{http://www.sciencedirect.com/science/article/pii/S0550321397800272}{{\em Nucl. Phys.} {\bf
  B492} (1997)  51--103}, \href{http://arxiv.org/abs/hep-ph/9610490}{{\tt
  arXiv:hep-ph/9610490}}.

\bibitem{Dine:1981gu}
M.~Dine and W.~Fischler, ``{A Phenomenological Model of Particle Physics Based
  on Supersymmetry}'',
\href{http://dx.doi.org/10.1016/0370-2693(82)91241-2}{{\em Phys. Lett.} {\bf
  B110} (1982)  227}.
%%CITATION = PHLTA,B110,227;%%.

\bibitem{AlvarezGaume:1981wy}
L.~Alvarez-Gaume, M.~Claudson, and M.~B. Wise, ``{Low-Energy Supersymmetry}'',
\href{http://dx.doi.org/10.1016/0550-3213(82)90138-9}{{\em Nucl. Phys.} {\bf
  B207} (1982)  96}.
%%CITATION = NUPHA,B207,96;%%.

\bibitem{Kaplan:1999ac}
D.~E. Kaplan, G.~D. Kribs, and M.~Schmaltz, ``{Supersymmetry breaking through
  transparent extra dimensions}'',
  \href{http://dx.doi.org/10.1103/PhysRevD.62.035010}{{\em Phys. Rev.} {\bf
  D62} (2000) no.~3, 035010},
\href{http://arxiv.org/abs/hep-ph/9911293}{{\tt arXiv:hep-ph/9911293
  [hep-ph]}}.
%%CITATION = HEP-PH/9911293;%%.

\bibitem{Chacko:1999mi}
Z.~Chacko, M.~A. Luty, A.~E. Nelson, and E.~Ponton, ``Gaugino mediated
  supersymmetry breaking'',
  \href{http://dx.doi.org/10.1088/1126-6708/2000/01/003}{{\em JHEP} {\bf 01}
  (2000)  003},
\href{http://arxiv.org/abs/hep-ph/9911323}{{\tt arXiv:hep-ph/9911323}}.
%%CITATION = HEP-PH 9911323;%%.

\bibitem{Ellis:1984bm}
J.~R. Ellis, C.~Kounnas, and D.~V. Nanopoulos, ``{No-Scale Supersymmetric
  GUTs}'',
\href{http://dx.doi.org/10.1016/0550-3213(84)90555-8}{{\em Nucl. Phys.} {\bf
  B247} (1984)  373--395}.
%%CITATION = NUPHA,B247,373;%%.

\bibitem{ourCascadePaper}
J.~Heisig and J.~Kersten, {\em {In preparation}}.

\end{thebibliography}\endgroup
\end{document}